# Human ≠ AGI

Roman V. Yampolskiy
Computer Science and Engineering
Speed School of Engineering
University of Louisville, USA
roman.yampolskiy@louisville.edu
July 6, 2020

*"A human being should be able to change a diaper, plan an invasion, butcher a hog, conn a ship, design a building, write a sonnet, balance accounts, build a wall, set a bone, comfort the dying, take orders, give orders, cooperate, act alone, solve equations, analyze a new problem, pitch manure, program a computer, cook a tasty meal, fight efficiently, die gallantly."*

- Robert A. Heinlein

*"There is no such thing as AGI.
There may be such a thing as human-level AI.
But human intelligence is nowhere near general."*

- Yann LeCun

**Abstract**
Terms *Artificial General Intelligence* (AGI) and *Human-Level Artificial Intelligence* (HLAI) have been used interchangeably to refer to the Holy Grail of Artificial Intelligence (AI) research, creation of a machine capable of achieving goals in a wide range of environments. However, widespread implicit assumption of equivalence between capabilities of AGI and HLAI appears to be unjustified, as humans are not general intelligences. In this paper, we will prove this distinction.

## 1. Introduction

Imagine that tomorrow a prominent technology company announces that they have successfully created an Artificial Intelligence (AI) and offers for you to test it out. You decide to start by testing developed AI for some very basic abilities such as multiplying 317 by 913, and memorizing your phone number. To your surprise, the system fails on both tasks. When you question the system's creators, you are told that their AI is human-level artificial intelligence (HLAI) and as most people cannot perform those tasks neither can their AI. In fact, you are told, many people can't even compute 13 x 17, or remember name of a person they just met, or recognize their coworker outside of the office, or name what they had for breakfast last

Tuesday[1]. The list of such limitations is quite significant and is the subject of study in the field of Artificial Stupidity [41, 40].

Terms *Artificial General Intelligence* (AGI) [13] and *Human-Level Artificial Intelligence* (HLAI) [6] have been used interchangeably (see [5], or "(AGI) is the hypothetical intelligence of a machine that has the capacity to understand or learn any intellectual task that a human being can." [1]) to refer to the Holy Grail of Artificial Intelligence (AI) research, creation of a machine capable of: achieving goals in a wide range of environments [23]. However, widespread implicit assumption of equivalence between capabilities of AGI and HLAI appears to be unjustified, as humans are not general intelligences. In this paper, we will prove this distinction.

Others use slightly different nomenclature with respect to general intelligence, but arrive at similar conclusions. "**Local generalization, or "robustness"**: … "adaptation to known unknowns within a single task or well-defined set of tasks". … **Broad generalization, or "flexibility":** "adaptation to unknown unknowns across a broad category of related tasks". …**Extreme generalization:** human-centric extreme generalization, which is the specific case where the scope considered is the space of tasks and domains that fit within the human experience. We … refer to "human-centric extreme generalization" as "generality". Importantly, as we deliberately define generality here by using human cognition as a reference frame …, it is only "general" in a limited sense. … To this list, we could, theoretically, add one more entry: "universality", which would extend "generality" beyond the scope of task domains relevant to humans, to any task that could be practically tackled within our universe (note that this is different from "any task at all" as understood in the assumptions of the No Free Lunch theorem [48, 47])." [10].

## 2. Prior work

We call some problems 'easy', because they come naturally to us like understanding speech or walking and we call other problems 'hard' like playing Go or violin, because those are not human universals and require a lot of talent and effort [50]. We ignore 'impossible' for humans to master domains, since we mostly don't even know about them or see them as important. As LeCun puts it: "[W]e can't imagine tasks that are outside of our comprehension, right, so we think, we think we are general, because we're general of all the things that we can apprehend, but there is a huge world out there of things that we have no idea" [22]. Others, agree: "we might not even be aware of the type of cognitive abilities we score poorly on." [4].

This is most obvious in how we test for intelligence. For example, Turing Test [42], by definition, doesn't test for universal general intelligence, only for human-level intelligence in human domains of expertise. Like a drunkard searching for his keys under the light because there it is easier to find them, we fall for the Streetlight effect observation bias only searching for intelligence in domains we can easily comprehend [56]. "The g factor, by definition, represents the single cognitive ability common to success across all intelligence tests, emerging from applying factor analysis to test results across a diversity of tests and individuals. But intelligence tests, by construction, only encompass tasks that humans can perform – tasks that are immediately recognizable and understandable by humans (anthropocentric bias), since including tasks that humans couldn't perform would be pointless. Further, psychometrics establishes measurement validity by demonstrating predictiveness with regard to activities that humans value (e.g. scholastic success): the very idea of a "valid" measure of intelligence only makes sense within the frame of reference of human values." [10].

---

[1]Some people could do that and more, for example 100,000 digits of $\pi$ have been memorized using special mnemonics.

Moravec further elaborates the difference between future machines and humans: "Computers are universal machines, their potential extends uniformly over a boundless expanse of tasks. Human potentials, on the other hand, are strong in areas long important for survival, but weak in things far removed. Imagine a "landscape of human competence," having lowlands with labels like "arithmetic" and "rote memorization," foothills like "theorem proving" and "chess playing," and high mountain peaks labeled "locomotion," "hand-eye coordination" and "social interaction." Advancing computer performance is like water slowly flooding the landscape. A half century ago it began to drown the lowlands, driving out human calculators and record clerks, but leaving most of us dry. Now the flood has reached the foothills, and our outposts there are contemplating retreat. We feel safe on our peaks, but, at the present rate, those too will be submerged within another half century." [31].

Chollet writes: "How general is human intelligence? The No Free Lunch theorem [48, 47] teaches us that any two optimization algorithms (including human intelligence) are equivalent when their performance is averaged across every possible problem, i.e. algorithms should be tailored to their target problem in order to achieve better-than-random performance. However, what is meant in this context by "every possible problem" refers to a uniform distribution over problem space; the distribution of tasks that would be practically relevant to our universe (which, due to its choice of laws of physics, is a specialized environment) would not fit this definition. Thus we may ask: is the human g factor universal? Would it generalize to every possible task in the universe? … [T]his question is highly relevant when it comes to AI: if there is such a thing as universal intelligence, and if human intelligence is an implementation of it, then this algorithm of universal intelligence should be the end goal of our field, and reverse-engineering the human brain could be the shortest path to reach it. It would make our field close-ended: a riddle to be solved. If, on the other hand, human intelligence is a broad but ad-hoc cognitive ability that generalizes to human-relevant tasks but not much else, this implies that AI is an open-ended, fundamentally anthropocentric pursuit, tied to a specific scope of applicability." [10].

Humans have general capability only in those human accessible domains and likewise artificial neural networks inspired by human brain architecture do unreasonably well in the same domains. Recent work by Tegmark et al. shows that deep neural networks would not perform as well in randomly generated domains as they do in those domains humans consider important, as they map well to physical properties of our universe. "We have shown that the success of deep and cheap (low-parameter-count) learning depends not only on mathematics but also on physics, which favors certain classes of exceptionally simple probability distributions that deep learning is uniquely suited to model. We argued that the success of shallow neural networks hinges on symmetry, locality, and polynomial log-probability in data from or inspired by the natural world, which favors sparse low-order polynomial Hamiltonians that can be efficiently approximated." [25].

## 3. Humans are not AGI

An agent is general (universal [19]) if it can learn anything another agent can learn. We can think of a true AGI agent as a superset of all possible NAIs (including capacity to solve AI-Complete problems [55]). Some agents have *limited domain generality*, meaning they are general, but not in all possible domains. The number of domains in which they are general may still be Dedekind-infinite, but it is a strict subset of domains in which AGI is capable of learning. For an AGI it's domain of performance is any efficiently learnable capability, while humans have a smaller subset of competence. Non-human animals in turn may have an even smaller repertoire of capabilities, but are nonetheless general in that subset. This means that humans can do things animals cannot and AGI will be able to do something no human can. If an AGI is restricted only to domains and capacity of human expertise, it is the same as HLAI.

Humans are also not all in the same set, as some are capable of greater generality (G factor [20]) and can succeed in domains, in which others cannot. For example, only a tiny subset of all people is able to conduct cutting-edge research in quantum physics, implying differences in our general capabilities between theory and practice. While theoretical definition of general intelligence is easy to understand, its practical implementation remains uncertain. "LeCun argues that even self-supervised learning and learnings from neurobiology won't be enough to achieve artificial general intelligence (AGI), or the hypothetical intelligence of a machine with the capacity to understand or learn from any task. That's because intelligence — even human intelligence — is very specialized, he says. "AGI does not exist — there is no such thing as general intelligence," said LeCun. "We can talk about rat-level intelligence, cat-level intelligence, dog-level intelligence, or human-level intelligence, but not artificial general intelligence."" [46].

An agent is not an AGI equivalent if it could not learn something another agent could learn. Hence, we can divide all possible tasks into human learnable and those, which no human can learn, establishing that humans are not AGI equivalent. We already described 'easy' and 'hard' for humans problems, the third category of 'impossible' is what we would classify as abilities impossible for humans to learn efficiently [43]. Computer-unaided humans [7] do not possess capabilities in this category, to any degree, and are unlikely to be able to learn them. If performed by a human, they would be considered magical, but as Arthur Clarke has famously stated: "Any sufficiently advanced technology is indistinguishable from magic."

Some current examples include: estimating face from speech [34], DNA [37] or ear [49], extracting passwords from typing sounds [38, 60], using lightbulbs [33] and hard drives [21] as microphones, communicating via heat emissions [16], or memory-write-generated electromagnetic signals [15], and predicting gender, age and smoking status from images of retinal fundus [35]. This is what is already possible with Narrow AI (NAI) today, AGI will be able to see patterns where humans see nothing but noise, invent technologies we never considered possible and discover laws of physics far above our understanding. Capabilities, we humans will never possess, because we are not general intelligences. Even humans armed with simple calculators are no match for such problems.

LeCun gives an example of one task no human could learn: "So let me take a very specific example, it's not an example it's more like a quasi-mathematical demonstration, so you have about 1 million fibers coming out of one of your eyes, okay two million total, but let's talk about just one of them. It's 1 million nerve fibers in your optical nerve, let's imagine that they are binary so they can be active or inactive, so the input to your visual cortex is 1 million bits. Now, they connected to your brain in a particular way and your brain has connections that are kind of a little bit like a convolution net they are kind of local, you know, in the space and things like this. Now imagine I play a trick on you, it's a pretty nasty trick I admit, I cut your optical nerve and I put a device that makes a random permutation of all the nerve fibers. So now what comes to your, to your brain, is a fixed but random permutation of all the pixels, there's no way in hell that your visual cortex, even if I do this to you in infancy, will actually learn vision to the same level of quality that you can." [22].

Chollet elaborates on the subject of human unlearnable tasks: "[H]uman intellect is not adapted for the large majority of conceivable tasks. This includes obvious categories of problems such as those requiring long-term planning beyond a few years, or requiring large working memory (e.g. multiplying 10-digit numbers). This also includes problems for which our innate cognitive priors are unadapted; … For instance, in the [Traveling Salesperson Problem] TSP, human performance degrades severely when inverting the goal from "finding the shortest path" to "finding the longest path" [29] – humans perform even worse in this case than one of the simplest possible heuristic: farthest neighbor construction. A particularly marked human bias is dimensional bias: humans … are effectively unable to handle 4D and higher. … Thus, … "general intelligence" is not a binary property which a system either possesses or lacks. It is a spectrum," [10]. "Human physical capabilities can thus be said to be "general", but only in a limited sense; when taking a broader view, humans reveal themselves to be extremely specialized, which is to be expected given the process through which they evolved." [10]. "[W]e are born with priors about ourselves, about the world, and about how to learn, which determine what categories of skills we can acquire and what categories of problems we can solve." [10].

If such tasks are in fact impossible for any human to perform, that proves that humans are not AGI equivalent. But, how do we know what a highly intelligent agent is capable of or more interestingly incapable of learning? How do we know what human's can't learn [61]? One trick we can use, is to estimate the processing speed [36] for an average human on a particular learning task and to show that even 120 years, a very optimistic longevity estimate for people, is not sufficient to complete learning that particular task, while much faster computer can do so in seconds.

Generality can be domain limited or unlimited. Different animals, such as dolphins, elephants, mice, etc. and humans are all general in overlapping but not identical sets of domains. Humans are not a superset of all animal intelligences. There are some things animals can do that humans cannot and vice versa. For example, humans can't learn to speak animal "languages" and animals can't learn to play chess [53]. Only AGI is universal/general intelligence over all learnable domains. AGI is not just capable of anything a human can do; it is capable of learning anything that could be learned. It is a Superset of all NAIs and is equal in capability to Superintelligence.

## 4. Conclusions

There is no shortage of definitions of intelligence [23, 24, 45, 18, 52], but we felt it was important to clarify that humans are neither fully general nor terminal point in the space of the possible minds [54]. As Chollet says: "We may even build systems with higher generalization power (as there is no a priori reason to assume human cognitive efficiency is an upper bound), or systems with a broader scope of application. Such systems would feature intelligence beyond that of humans." [10]. Humans only have a subset of capabilities an AGI will have and the capability difference between us and AGI is far greater than capability difference between AGI and superintelligence (SAI). Bostrom describes three forms of superintelligence (p. 53-57) [8]): Speed SAI (like a faster human), Collective SAI (like a group of humans), and Quality SAI (does what humans can't). All three can be accomplished by an AGI, so there is no difference between AGI and SAI, they are the same (HLAI ≤ AGI = SAI) and the common takeoff-speed debate [59] resolves to hard takeoff, from definitions. This implies even stronger limitations [57, 56, 58] on our capability to control AI and a more immediate faceoff. We are already having many problems with Ignorance Explosion [27, 28], an Intelligence Explosion [32, 26] will be well beyond our capabilities to control.

If we use Legg's definition of intelligence [23], and average performance across all possible problems, we can arrive at a somewhat controversial result that modern AI is already smarter than any human is. An individual human can only learn a small subset of domains and human capabilities can't be trivially transferred between different humans to create a union function of all human capabilities, but that is, at least theoretically, possible for AI. Likewise, humans can't emulate some computer algorithms, but computers can run any algorithm a human is using. Machines of 2020 can translate between hundreds of languages, win most games, generate art, write poetry and learn many tasks individual humans are not capable of learning. If we were to integrate all such abilities into a single AI agent it would on average outperform any person across all possible problem domains, but perhaps not humanity as a whole seen as a single agent. This may have been true for a number of years now, and is becoming more definitive every year. As an AI agent can be a superset of many algorithms from which it can choose it would not be a subject to the No Free Lunch (NFL) theorems [48, 47].

While AI dominates humans in most domains of human interest [12, 14, 30, 39, 11, 44], there are domains in which humans would not even be able to meaningfully participate. This is similar to the Unpredictability [57] and Unexplainability/Incomprehensibility of AI [56] results, but at a meta-level. The implications for AI control and AI Safety and Security [51, 9, 2, 3] are not encouraging. To be dangerous AI doesn't have to be general, it is sufficient for it to be superior to humans in a few strategic domains. If AI can learn a particular domain it will quickly go from Hypohuman to Hyperhuman performance [17]. Additionally, common proposal for merging of humanity with machines doesn't seem to work as adding HLAI to AGI adds nothing to AGI, meaning in a cyborg agent human will become a useless bottleneck as AI becomes more advanced and the human will be eventually removed, if not explicitly at least implicitly from control. What does this paper tell us? Like the dark matter of the physical universe, the space of all problems is mostly unknown unknowns, and most people don't know that and don't even know that they don't know it. To paraphrase the famous saying: "The more AI learns, the more I realize how much I don't know."

## Acknowledgments
The author is grateful to Elon Musk and the Future of Life Institute, and to Jaan Tallinn and Effective Altruism Ventures for partially funding his work on AI Safety. Author regrets not being a general intelligence, as that would greatly improve quality of his papers.

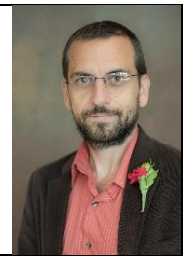

Dr. Roman V. Yampolskiy is Tenured Faculty in the department of Computer Science and Engineering at the University of Louisville. He is the founding and current director of the Cyber Security Lab and an author of multiple books including Artificial Superintelligence: a Futuristic Approach. Dr. Yampolskiy's main areas of interest are Artificial Intelligence Safety and Security. Twitter: @romanyam